# A numerical study of the effect of discretization methods on the crystal plasticity finite element method

Jingwei Chen, Zifan Wang, Alexander M. Korsunsky *

*MBLEM, Department of Engineering Science, University of Oxford, Parks Road, Oxford OX1 3PJ, United Kingdom*

jingwei.chen@eng.ox.ac.uk

zifan.wang@exeter.ox.ac.uk

alexander.korsunsky@eng.ox.ac.uk, *corresponding author

## Abstract

The present report describes a big data numerical study of crystal plasticity finite element (CPFE) modelling using static and grain-based meshing to investigate the dependence of the results on the discretization approach. Static mesh refers to the integration point-based representation of the microstructure in which the integration points (IPs) within a finite element may belong to different grains, while in the grain-based meshing the minimum discretization unit is an element that may only belong to one grain. The crystal plasticity constitutive law was coded using UMAT subroutine within commercial finite element software Abaqus. Multiple sets of RVEs were investigated under strain-controlled loading and periodic boundary conditions. The stress and strain contour maps obtained from RVEs with static mesh and grain-based mesh were compared. The simulation results reveal that both discretization methods provide reliable predictions of the stress-strain curves and the stress/strain localization points in polycrystalline alloys. Static mesh tends to smooth the stress/strain profile at the grain boundary, whilst stress/strain discontinuities are present in the grain-based mesh results. The above findings remain valid when the number of grains within an RVE increases from 34 to 1250. To quantify the difference between static and grain-based meshing, a relative measure of deviation is defined. The deviations of global stress were found to be relatively small, within 0.5%, while local deviations were significant up to 50%. Static mesh has the advantage of reducing both the pre-processing procedures and computational time compared to grain-based mesh. It is concluded that static mesh is preferred when investigating the material's *macroscopic* behaviour, whilst grain-based mesh is recommended for the study of the *local* response using CPFEM.

## Keywords

# 1. Introduction

Until the 1980s the prediction of the deformation fields within materials relied mainly on continuum level models, neglecting their inhomogeneous nature related to material microstructure. It has since become clear from both experimental and modelling studies that such a description is not reliable when the characteristic length of the investigation reduces to the scale of micrometres. For polycrystalline metals such as nickel-based superalloys, the elastic and plastic deformation are highly anisotropic. Plastic deformation occurs mainly by the movement of dislocation along specific slip planes and directions. Depending on their crystallographic orientation with respect to the loading direction, grains deform to different extents to carry disparate magnitudes of stress under the applied load, giving rise to heterogeneous deformation fields between adjacent grains and within individual grains [1].

With the advances in computational mechanics in recent decades, different crystal plasticity frameworks were established and applied to predict the deformation behaviour of single crystals and polycrystals. The implementation of crystal plasticity theory within the finite element method is called the crystal plasticity finite element method (CPFEM). CPFEM studies appear to have been initiated in the single crystal multiple slip study by Peirce et al. [2] and have been extensively used since to investigate the heterogeneous deformation response of polycrystals under various conditions [3-7]. The crystal orientation and material state variables are updated in each load increment at every integration point (IP) by solving the crystal plasticity formulation. CPFEM not only predicts the macroscopic stress-strain curve and texture evolution, but also provides excellent predictions of the local deformation fields and local crystal orientation with the help of appropriate homogenization schemes [8-15], such as a representative volume element (RVE).

Based on the discretization methods, the implementation of CPFEM can be categorized into two types: (i) those considering crystal plasticity constitutive model at the integration point level (called static mesh in this paper); (ii) those representing the grain morphology at the element level (called grain-based mesh). For static mesh, it is integration point-based and the IPs in the same element may belong to different grains. Before 2010s, static mesh was widely used to investigate the microstructure-sensitive material properties, i.e., texture evolution [16,18], stress concentration [5], phase transformation [17], fatigue criteria [19,20]. This discretization approach continues to be employed by researchers nowadays to study the mechanical properties of various alloys including FeCrAl alloys [21], ultrafine-grained nickel [22] and steel [23]. Grain-based meshing that became more popular with the increase in computational power uses a finite element as the minimum discretization unit, so that IPs within the same element may only belong to one grain. Over the past decade, the grain-based mesh became favoured over the static mesh method due to its advantage in visualization and post-processing [24-30]. Although a number of researchers employ both static and grain-based meshing in CPFEM, no published data provide a comprehensive and systematic comparative study of the difference that arises between the two discretization approaches.

The present study aims to investigate the effect of discretization methods and grain numbers within an RVE for CPFEM. To represent complex grain interaction scenarios in realistic microstructures, the present investigation considers two element types including static and grain-based mesh of various grain numbers that range from 34 to 1250 in predicting the stress and strain field. Material properties and grain orientations were assigned at integration points for static mesh and elements for grain-based mesh. Multiple RVE realizations with different grain numbers were generated by Voronoi tessellation and massive simulations were performed to reveal the effect of discretization methods and grain numbers. Periodic boundary conditions and the same number of elements per grain were employed for all the RVE realizations to aid precise comparison. CPFEM presented are analysed in terms of strain/stress fields accuracy and computational efficiency. The conclusions of this study provide further insights into the rational choice of discretization methods in polycrystalline CPFE modelling.

## 2. Modelling methodology

### 2.1 Crystal plasticity formulation and the hardening law

Some researchers studied the effect of crystal plasticity hardening framework on the mechanical response of polycrystalline alloys. Lim et al. suggested that slip-based hardening law can accurately reproduce the deformation behaviour obtained from dislocation density-based constitutive equations

[31]. Additionally, slip-based hardening law shows smaller relative error and smaller mesh sensitivity compared to dislocation density-based law. Therefore, phenomenological constitutive (slip-based) hardening law is chosen in this research. The present model was implemented using phenomenological crystal plasticity constitutive equations following Manonukul & Dunne [32]. It was further extended to account for elastic anisotropy, and to allow three-dimensional modelling for alloys with various crystal structures, i.e. face-centred cubic (FCC), body-centred cubic (BCC) and hexagonal-close packed (HCP) crystal lattice types.

The crystal plasticity equation is based on the multiplicative decomposition of the deformation gradient $\boldsymbol{F}$ into elastic and plastic parts

$$\boldsymbol{F} = \boldsymbol{F}^E \boldsymbol{F}^P \tag{1}$$

The material undergoes plastic slip $\gamma^\alpha$ on the slip system $\alpha$, through the undeformed crystal lattice, the plastic deformation gradient can be expressed as

$$\boldsymbol{F}^P = \boldsymbol{I} + (\boldsymbol{s}^\alpha \boldsymbol{n}^{\alpha T})\gamma^\alpha \tag{2}$$

Here $\boldsymbol{I}$ is the identity tensor, $\boldsymbol{s}^\alpha$ and $\boldsymbol{n}^\alpha$ are two mutually orthogonal vectors describing the slip direction and normal, respectively. The term $\boldsymbol{F}^E$ represents both elastic deformation and rigid-body rotation. The stress rate at arbitrary locations during the slip process is described as

$$\dot{\boldsymbol{\sigma}} = \boldsymbol{C}:\boldsymbol{D} - \boldsymbol{\sigma}\mathrm{tr}(\boldsymbol{D}) - \boldsymbol{\Omega}\boldsymbol{\sigma} + \boldsymbol{\sigma}\boldsymbol{\Omega} - \sum_{\alpha=1}^{N}(\boldsymbol{C}:\boldsymbol{P}^\alpha + \beta^\alpha)\dot{\gamma}^\alpha \tag{3}$$

Here $\boldsymbol{C}$ is the elastic moduli tensor, $\boldsymbol{D}$ represents the deformation rate, and $\boldsymbol{\Omega}$ describes the spin tensor. $\beta^\alpha = \boldsymbol{W}^\alpha\boldsymbol{\sigma} - \boldsymbol{\sigma}\boldsymbol{W}^\alpha$, $\boldsymbol{P}^\alpha$ is the symmetric part of the velocity gradient, and $\boldsymbol{W}^\alpha$ is the skew part of the velocity gradient. The overall plastic strain rate comes from the contribution of all active slip systems. The constitutive model uses a critical resolved shear stress, $\tau_c^\alpha$, as a state variable for the determination of plastic flow on each slip system $\alpha$. When the resolved shear stress on the $\alpha$th slip system, $\tau^\alpha$, exceeds the current critical resolved shear stress $\tau_c^\alpha$, that slip system is considered to become active. The critical resolved shear stress reflects the resistance of a slip system, and the slip resistance originates the material properties such as current dislocation density and substructure. The conditions under which a slip system is active are based on the yield and loading-unloading criteria. The shearing rate $\dot{\gamma}^\alpha$ on the $\alpha$th active slip system can be determined using the constitutive equation for plastic slip on each slip system,

$$\dot{\tau}_c^\alpha = \sum_{\alpha=1}^{N} h_{\alpha\beta}\, \dot{\gamma}^\alpha \qquad (4)$$

Here $\tau_c^\alpha$ is the current critical resolved shear stress. The hardening modulus $h_{\alpha\beta}$ represents the hardening on the slip system $\alpha$ due to shearing on the slip system $\beta$. The rate of change of the resolved shear stress is calculated by

$$\dot{\tau}^\alpha = P^\alpha:\left(C:D - C:\sum_{\alpha=1}^{N} P^\alpha\, \dot{\gamma}^\alpha - \sigma\mathrm{tr}(D)\right) + \beta^\alpha:\left(D - \sum_{\alpha=1}^{N} P^\alpha\, \dot{\gamma}^\alpha\right) \qquad (5)$$

To satisfy the consistency condition during plastic slip,

$$\dot{\tau}^\alpha = \dot{\tau}_c^\alpha \qquad (6)$$

Therefore, from equations (4) and (5),

$$\sum_{\alpha=1}^{N} h_{\alpha\beta}\, \dot{\gamma}^\alpha = P^\alpha:\left(C:D - C:\sum_{\alpha=1}^{N} P^\alpha\, \dot{\gamma}^\alpha - \sigma\mathrm{tr}(D)\right) + \beta^\alpha:\left(D - \sum_{\alpha=1}^{N} P^\alpha\, \dot{\gamma}^\alpha\right) \qquad (7)$$

Equation (7) can be rewritten in the form

$$\sum_{\alpha=1}^{N} A_{\alpha\beta}\, \dot{\gamma}^\alpha = b^\alpha \qquad (8)$$

When the matrix $A_{\alpha\beta}$ is singular, the above equation has non-unique solutions. This problem can be solved by using a singular-value decomposition to obtain a pseudo-inverse of the matrix $A_{\alpha\beta}$ [32], and it is adopted here. The hardening modulus matrix can be simply described as

$$h_{\alpha\beta} = [q + (1-q)\delta_{\alpha\beta}]h. \qquad (9)$$

Here $q$ is the latent-hardening rate and $h$ is the self-hardening ratio. The value of $q$ ranges from 1 to 1.4. The above crystal plasticity constitutive equations are implemented using a user-defined material subroutine UMAT within the commercial finite element analysis package ABAQUS.

## 2.2 Microstructure generation procedures

For static mesh, the RVE of the polycrystal is discretized as a fixed hexahedral mesh (C3D8 element) and the material properties are assigned to each **IP**. The open-source software package Neper [33] is utilized to generate the grain origin positions ('seeds') that are associated with the crystal orientation specified at each grain. To reflect the grain growth during the crystallization process, a population of 34 grains is derived from the seeds by 3D Voronoi tessellation. By varying the spatial distribution of the seeds, it is possible to control grain size distribution within RVE. Some published CPFEM studies suggest that 50-100 elements per grain are sufficient to predict microscopic deformation field [34-36]. Thus, 15x15x15 elements are employed in this study (as shown in Figure 1(a)), leading to an average of ~100 elements per grain. In contrast with grain-based meshing, this approach uses a static FE mesh in combination with a dynamic assignment of crystal orientation per IP, allowing a flexible implementation of both grain morphology and orientation, jointly or severally. The use of integration-point-based assignment of grain orientation will improve the accuracy and efficiency of diffraction post-processing when calculating grain orientation-average elastic strain from CPFEM [34]. The initial crystal orientation is assigned at each IP by inheriting it from the nearest grain seed. Texture can be described by the choice of grain orientation distribution function (ODF). In this study, a set of RVE together with the assigned initial crystal orientations are referred to as a microstructural realization. The main advantage of static mesh is that it can be used to investigate statistical material response induced by several realizations without changing the RVE model once the deformation model is validated.

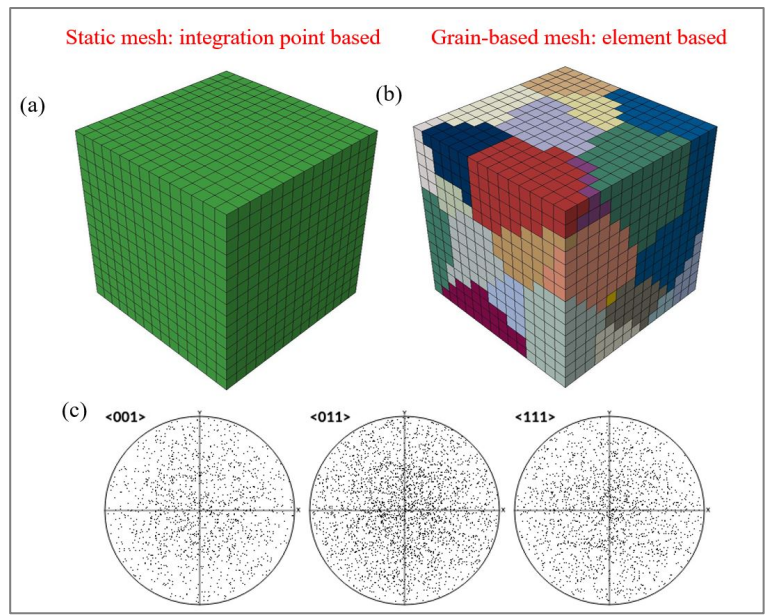

**Figure 1**. The RVE models for (a) static mesh and (b) grain-based mesh. (c) pole figures show random texture in three directions. The same tessellated microstructure is used for both static and grain-based mesh. Different colours represent different grains.

For grain-based mesh, the RVE of the polycrystal is also discretized as C3D8 elements while the material properties are assigned to each **element**. The same set of seeds and corresponding crystal orientation are employed to generate a grain-based RVE by Neper software, as illustrated in Figure

1(b). The set of elements that belong to the same grain is defined in the pre-processing stage, and hence, the RVE model needs to be changed when investigating the mechanical property of different microstructural realizations. It is more convenient to visualize the grain structure for grain-based mesh compared to static mesh. The same lognormal grain size distribution and random ODF (shown in Figure 1 (c)) are utilized to construct the RVEs with static mesh and grain-based mesh throughout the study.

**2.3 Loading and boundary conditions**

In polycrystalline alloys, the deformation of RVE is inhomogeneous due to the nature of material anisotropy. To eliminate the effect of the boundary layer, periodic boundary conditions (PBC) instead of homogeneous boundary was applied to the initial configuration of RVE. Figure 2 depicts the initial configuration of the RVE subjected to periodic boundary conditions. For one arbitrary node on the surface of RVE, there is a matching node on the opposite RVE boundary. PBC is implemented by coupling the displacements of the nodes of each pair of opposites faces of the RVE. Any over-constraint should be avoided for the nodes that are located on the vertices and edges. The displacement constraint equations were generated by the open-source ABAQUS plugin EasyPBC [37]. After setting the constraint equations, displacement boundary conditions were applied at RVE vertexes. A displacement-controlled loading was applied at the dummy node that constrains the relative displacement between the front and back surfaces in the third direction. Internal length scales are not considered in the study and all the RVEs have the same volume of $1\times1\times1\text{mm}^3$.

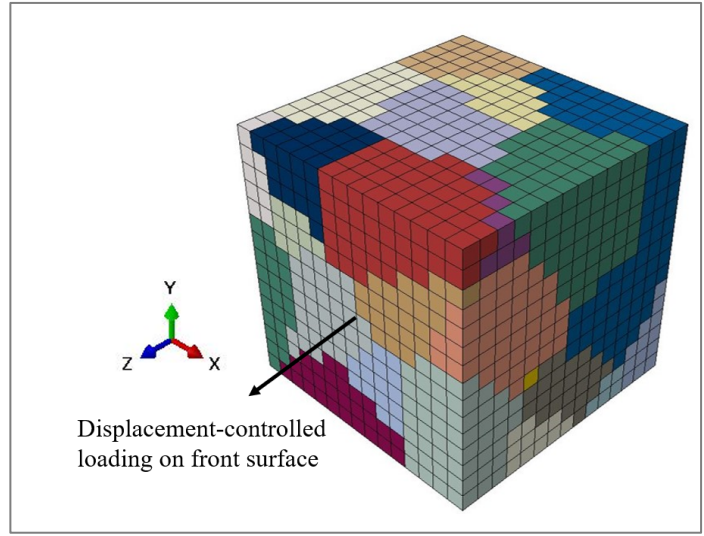

**Figure 2.** The RVE realization is subjected to periodic boundary conditions.

## 2.4 Model calibration and material parameters

Four material parameters need to be calibrated in this model: the initial critical resolved shear stress $\tau_c^\alpha$, two hardening parameters, and the elastic moduli matrix. To calibrate these parameters, the overall response of an RVE with 600 grains during tensile loading was fitted to the experimental stress-strain curves for a nickel-based superalloy, Haynes 282. It has an FCC crystal structure with 12 $\langle 1\bar{1}0\rangle\{111\}$ slip systems. Optimal values of materials were determined by fitting the macroscopic and mesoscopic response of RVE to the experimental stress-strain curve and neutron diffraction measurement obtained by Jaladurgam et al. [38]. For a detailed description of the calibration procedure, readers can refer to our previous publications [34,39]. Figure 3 compares the predicted stress-strain curve with the experimental curve and an excellent agreement has been achieved. The corresponding material parameters are listed in Table 1.

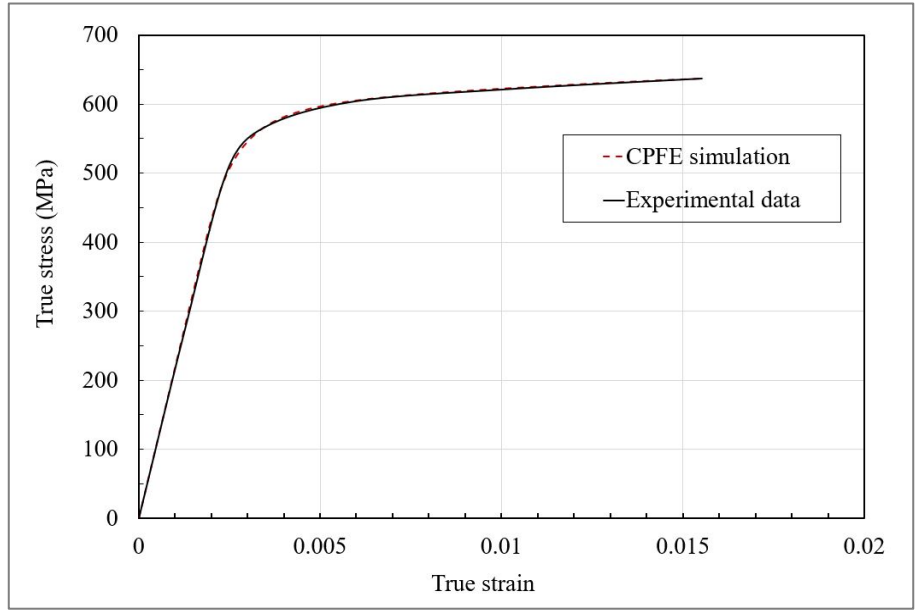

**Figure 3**. The macroscopic stress-strain curves under monotonic loading obtained from CPFE prediction and experiment.

Table 1. Material parameters used in the CPFEM simulation

| Stiffness (GPa) | | | CRSS (MPa) | h | q |
|---|---|---|---|---|---|
| C11 | C12 | C44 | | | |
| 250 | 160 | 118 | 257 | $h=h_F[1 + (h_R - 1)exp(-100h_{exp}\varepsilon)]$ * | 1.01 |

*$h_F = 950, h_R = 14, h_{exp} = 4$.

## 3. Simulation results and discussion

### 3.1 The effect of discretization methods

In the first part of the study, we concentrate on the difference in mechanical response caused by different discretization methods. Displacement-controlled tensile loading was applied to both RVEs with 34 grains shown in Figure 1. The RVEs with the same tessellated microstructure were discretized into the static and grain-based mesh, respectively. Both simulations were stopped when the total strain reached 1.55% and the overall stress of the whole RVE at 1.55% strain was termed as $\sigma_{1.55\%}$. Figure 4 depicts the macroscopic stress-strain curve obtained from RVEs with static and grain-based mesh. The two curves are almost identical with a minor discrepancy in the plastic region, which reveals that different discretization methods introduce little difference in terms of macroscopic response.

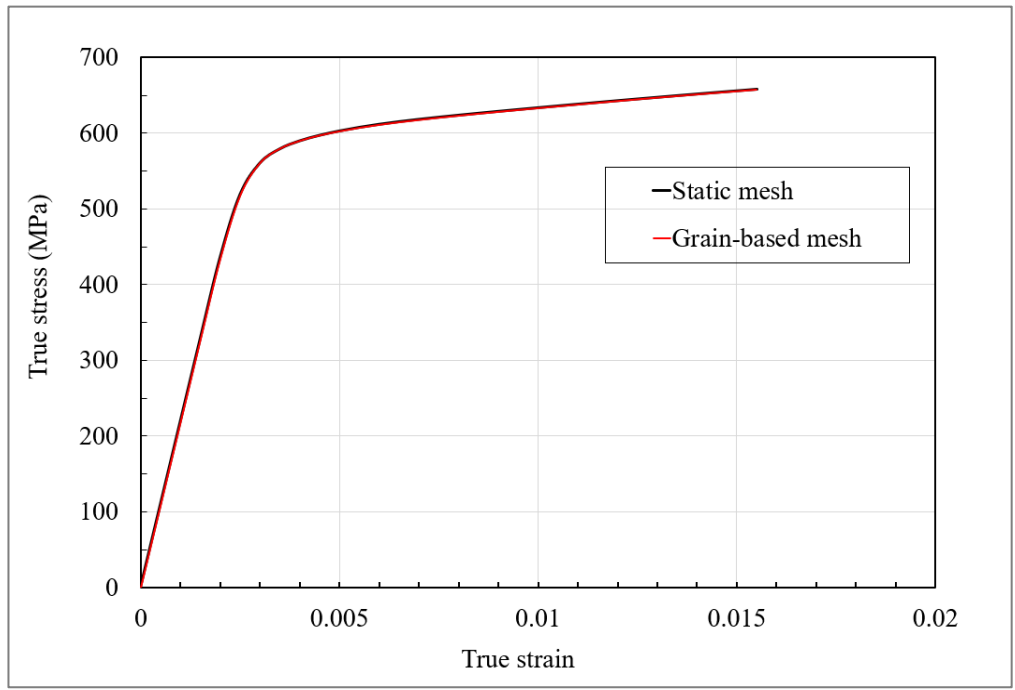

**Figure 4**. Illustration of stress-strain curves predicted from CPFEM with static and grain-based mesh.

Figures 5 and 6 illustrate the stress in the loading direction $\sigma_{33}$ and accumulated plastic strain after uniaxial tension to a strain of 1.55%. To further study the internal stress and strain field, the original RVE configuration was cut in half as shown in Figures 5 and 6. The apparent sensitivity of the models to discretization methods in terms of local (microscopic) mechanical response can be observed. A qualitative comparison of contour maps suggests that the CPFEM models can well predict stress/strain localization points and weak/cold locations in the polycrystalline materials independent of the two discretization methods investigated in this research. However, the predicted stress/strain values show obvious deviations at both inner and surface layers. Further investigation reveals that the largest heterogeneities and strain/stress discontinuities are present in the RVE with grain-based mesh.

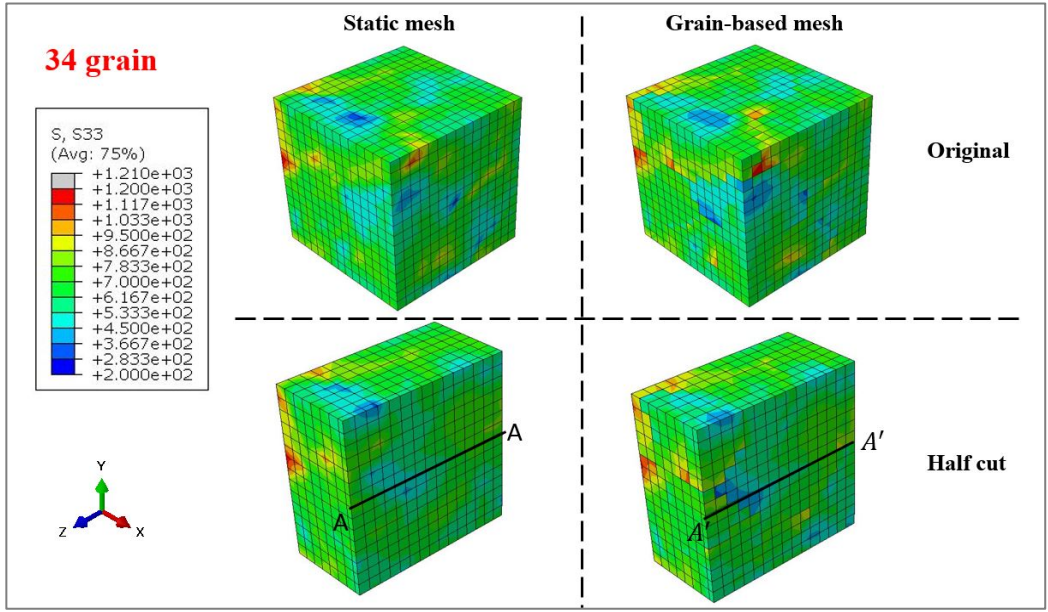

**Figure 5**. Original and half-cut contour maps show the stress in the loading direction after tension to a total strain of 1.55%.

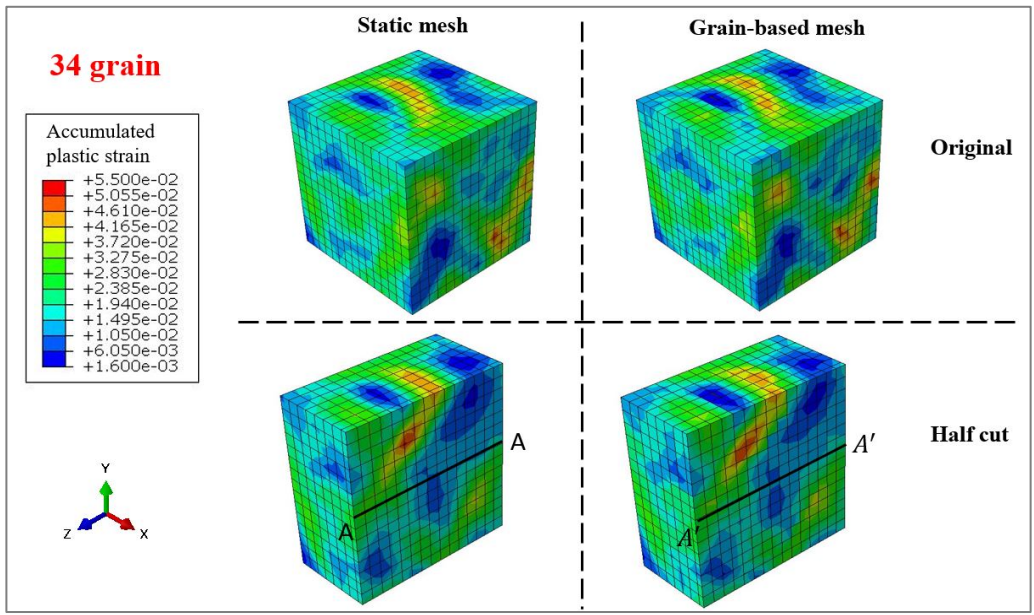

**Figure 6**. Original and half-cut contour maps show the accumulated plastic strain after tension to a total strain of 1.55%.

To investigate the origin of stress and strain discontinuities, path A-A and path $A'$-$A'$ were drawn at the same location in half-cut RVEs with static and grain-based mesh, respectively. The stress $\sigma_{33}$ and accumulated plastic strain across the two paths were shown in Figures 7 (a) and (b). The strain and stress profiles predicted by static mesh and grain-based mesh show the same trend and comparable magnitude. Stress/strain discontinuities are mainly observed at grain boundaries, and more discontinuities are found for the RVE with grain-based mesh. Two grain boundaries are located at the blue dash-dotted lines in Figure 7(a). The stress in the loading direction is continuous for static mesh, while discontinuous for grain-based mesh. Stress discontinuity often origins from material discontinuity. Here, grains with various orientations have different mechanical properties and can sustain different magnitudes of stress due to crystal anisotropy. Further investigation suggests that when the element of interest belongs to a single grain at the material discontinuity or grain boundary, stress discontinuity will occur at that discontinuity location for both static mesh and grain-based mesh. Nevertheless, when one element belongs to two or more grains at the grain boundary for static mesh, the stress discontinuity will be observed at the discontinuity location only for grain-based mesh, not for static mesh. The above finding indicates that static mesh will tend to reduce stress discontinuity and smooth stress profile at the material discontinuous location compared with grain-based mesh.

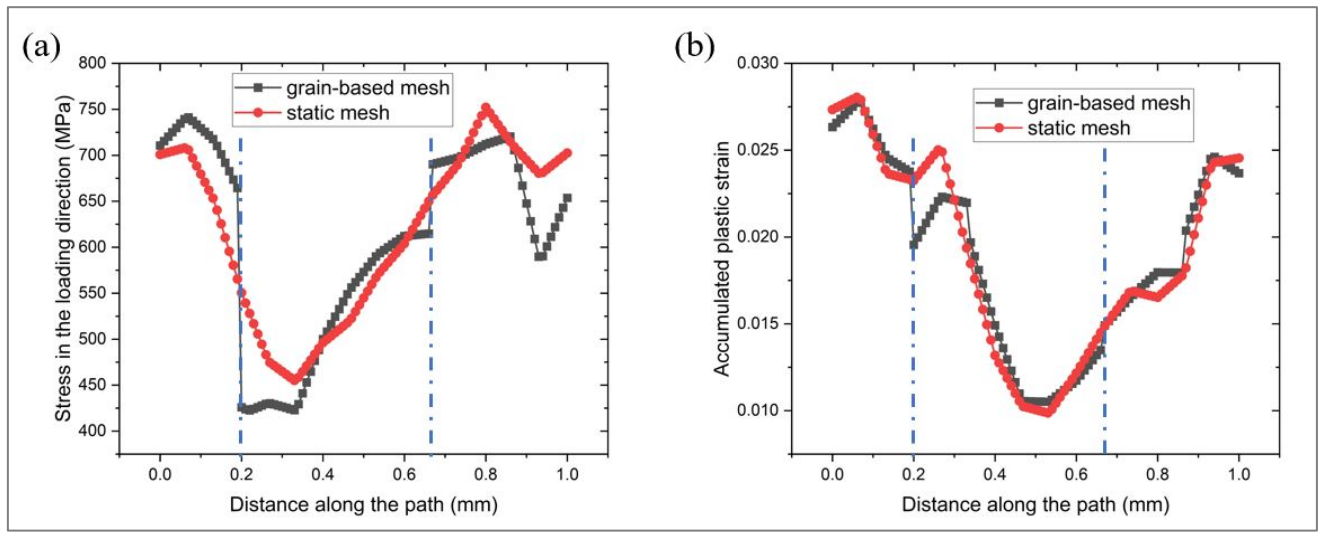

**Figure 7**. The stress in the loading direction and accumulated plastic strain across paths A-A and path $A'$-$A'$.

## 3.2 The effect of discretization methods with various grain numbers

The size effect in the CPFEM can be divided into discretization induced size effect and statistically induced size effect. The former effect arises from the discretization of grain (the number of elements per grain), while the latter is caused by different numbers of grains in the RVE. It is important to eliminate the discretization induced size effect when investigating the effect of discretization methods with various grain numbers. Therefore, the number of elements per grain is the same for all RVEs in this study. A comprehensive review of the literature shows that most of the CPFEM studies employ ~20-1500 grains in the RVE. As illustrated in Figure 8, four sets of RVEs with four grain numbers were generated in this study to represent most of the simulations in the literature. Each set of RVEs was discretized into the static mesh and grain-based mesh, and the same random orientation distribution function was used for all sets of RVE. Table 2 lists the realization numbers together with the assigned grain numbers and elements.

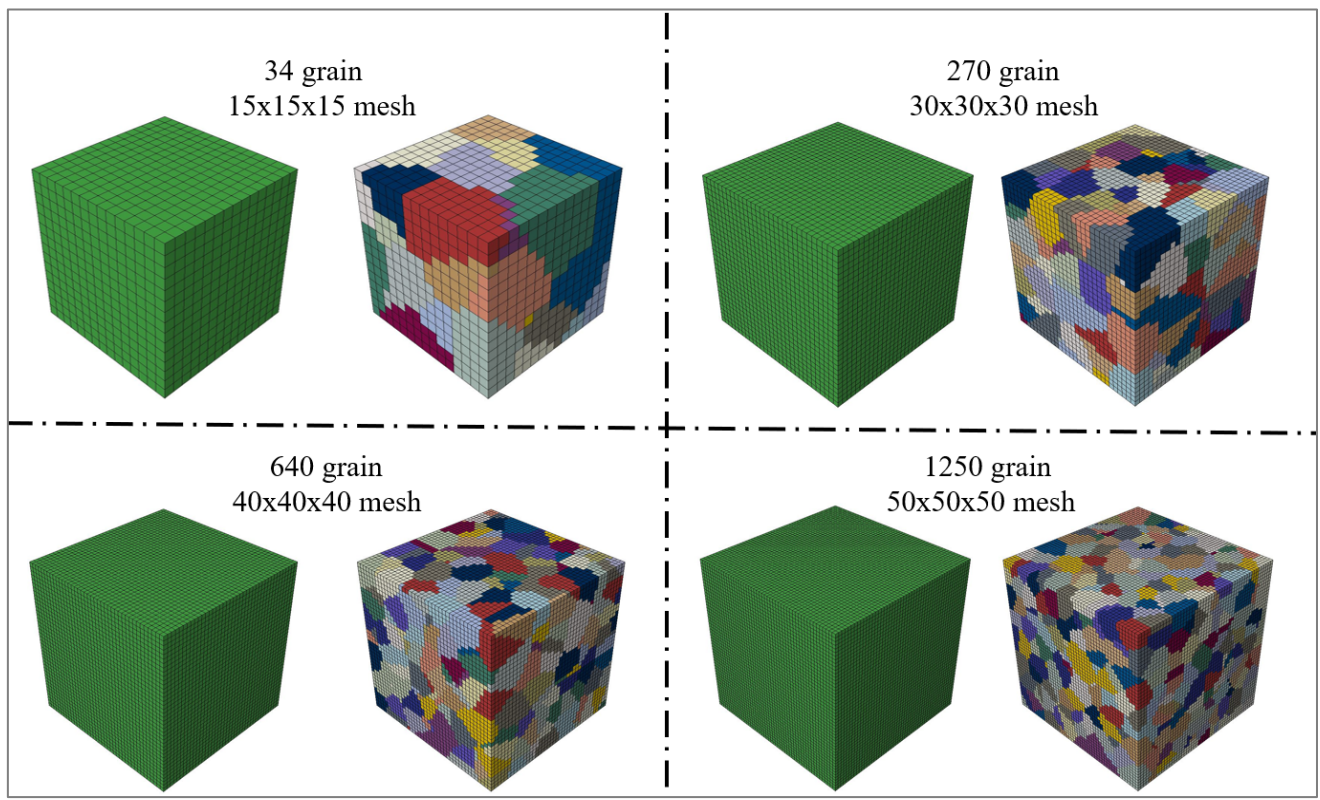

**Figure 8.** Four sets of RVE with different grain numbers and different discretization methods.

**Table 2**. Sets of realizations generated in the current research.

| Realization | No. of grains in RVE | Discretization method | Total number of elements | The average number of elements per grain |
|---|---|---|---|---|
| 1 | 34 | static | 3375 | ~100 |
| 2 | 34 | grain-based | 3375 | ~100 |
| 3 | 270 | static | 27000 | 100 |
| 4 | 270 | grain-based | 27000 | 100 |
| 5 | 640 | static | 64000 | 100 |
| 6 | 640 | grain-based | 64000 | 100 |
| 7 | 1250 | static | 125000 | 100 |
| 8 | 1250 | grain-based | 125000 | 100 |

The same displacement-controlled loading was applied to all realizations, and the predicted stress-strain curves are shown in Figure 9. Similar stress-strain curves are observed independent of discretization methods and grain numbers. An enlarged view in Figure 9 (b) suggests that the maximum difference between the curves obtained from static mesh and the corresponding grain-based mesh is less than 0.5%. Figures 10-15 illustrate the original and half-cut contour maps for RVE with 270,640 and 1250 grains after tension to a total strain of 1.55%. Analogous to the findings for RVEs with a smaller number of grains as shown in Figures 5 and 6, the results for RVEs with a larger number of grains also reveal that the CPFEM models can predict stress/strain localization position and weak/cold spots independent of discretization methods. The magnitude of predicted local strain/stress shows apparent deviation and more strain/stress discontinuities are observed in the RVE with grain-based mesh.

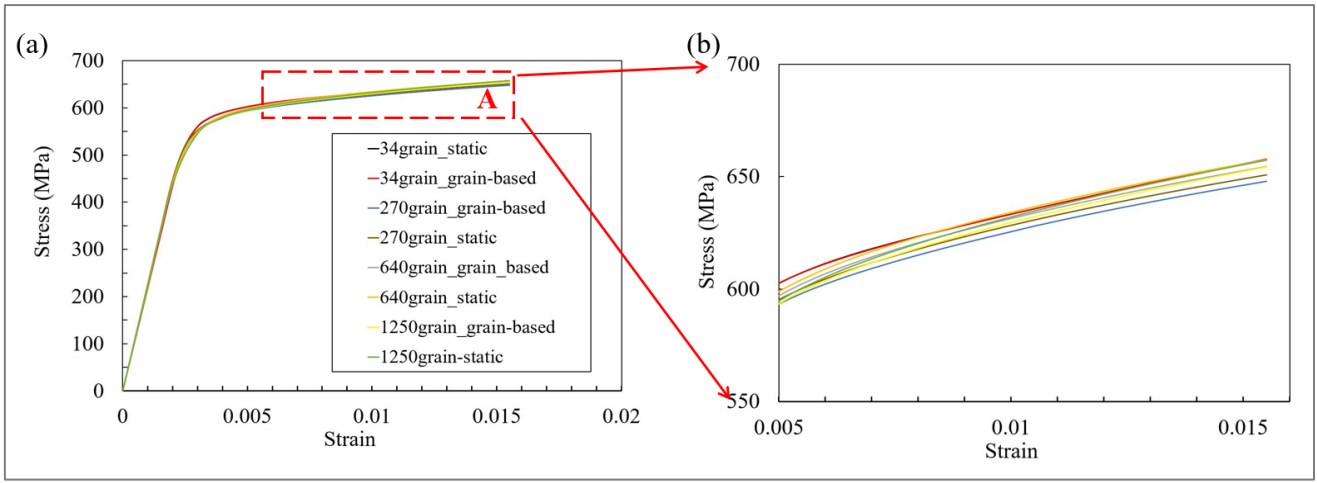

**Figure 9**. (a) The stress-strain curves from different discretization methods and grain numbers. (b) An enlarged view of area A.

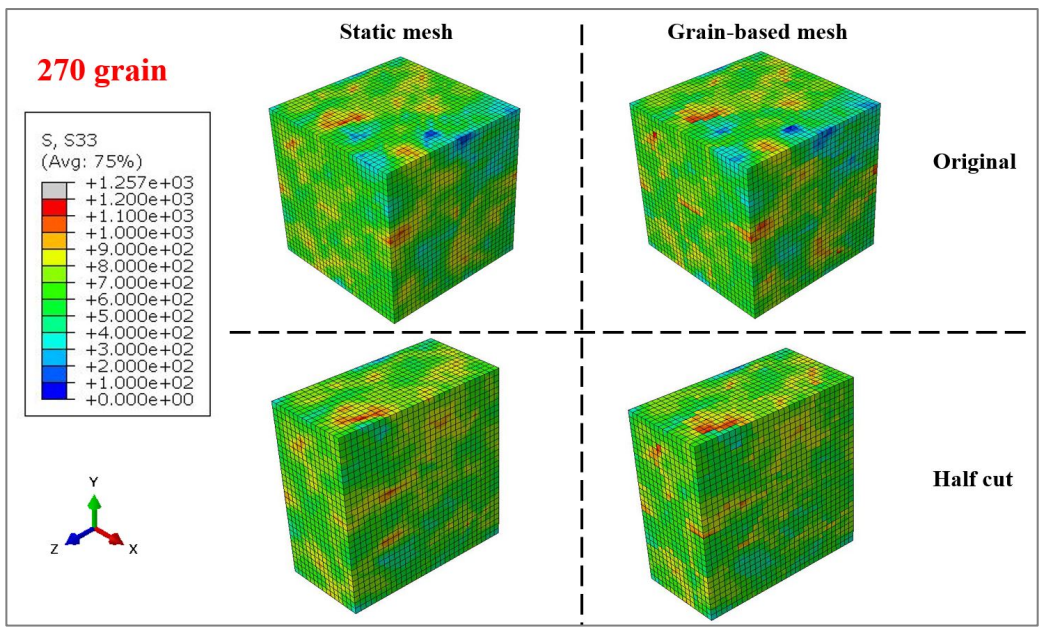

**Figure 10**. Original and half-cut contour maps for RVE with 270 grains show the stress in the loading direction after tension to a total strain of 1.55%.

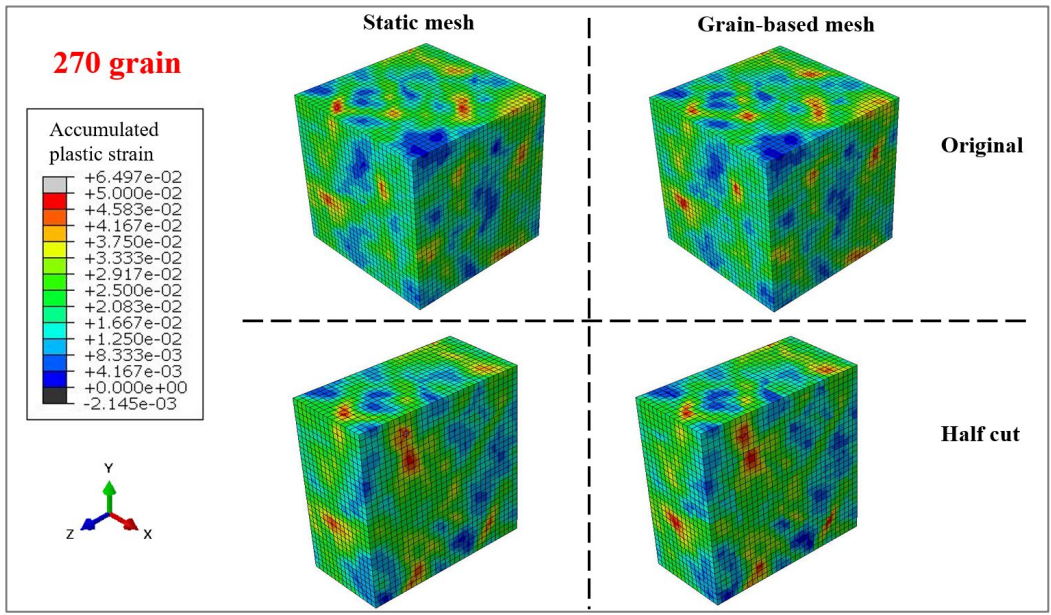

**Figure 11**. Original and half-cut contour maps for RVE with 270 grains show accumulated plastic strain after tension to a total strain of 1.55%.

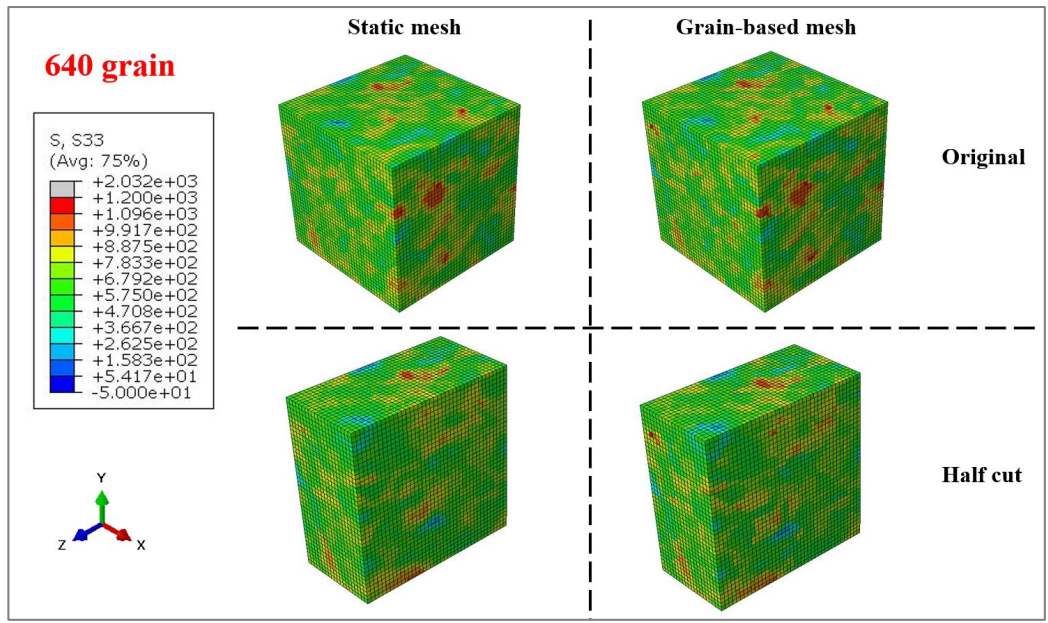

**Figure 12**. Original and half-cut contour maps for RVE with 640 grains show the stress in the loading direction after tension to a total strain of 1.55%.

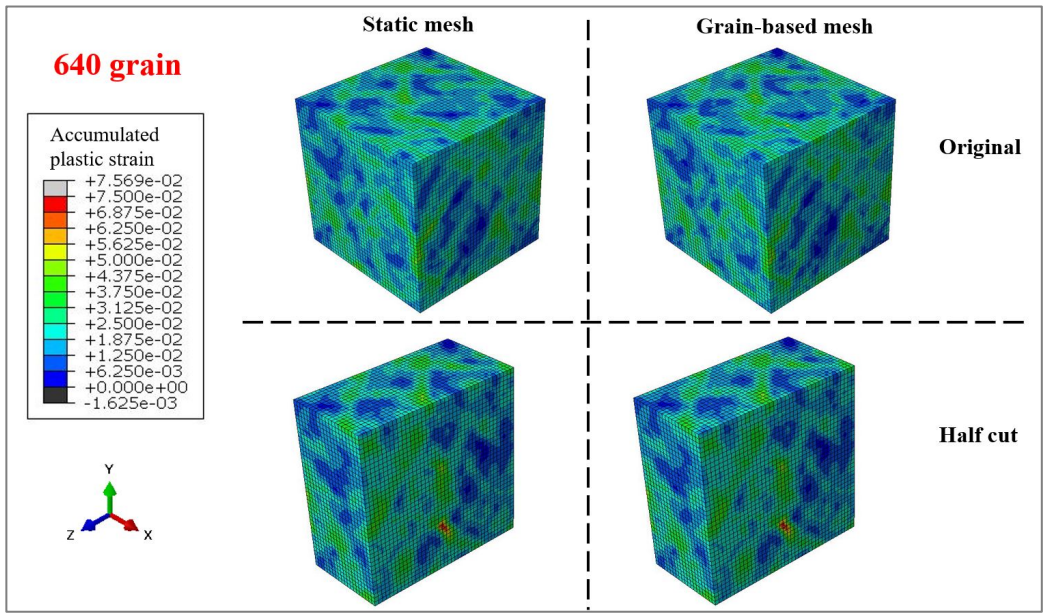

**Figure 13**. Original and half-cut contour maps for RVE with 640 grains show accumulated plastic strain after tension to a total strain of 1.55%.

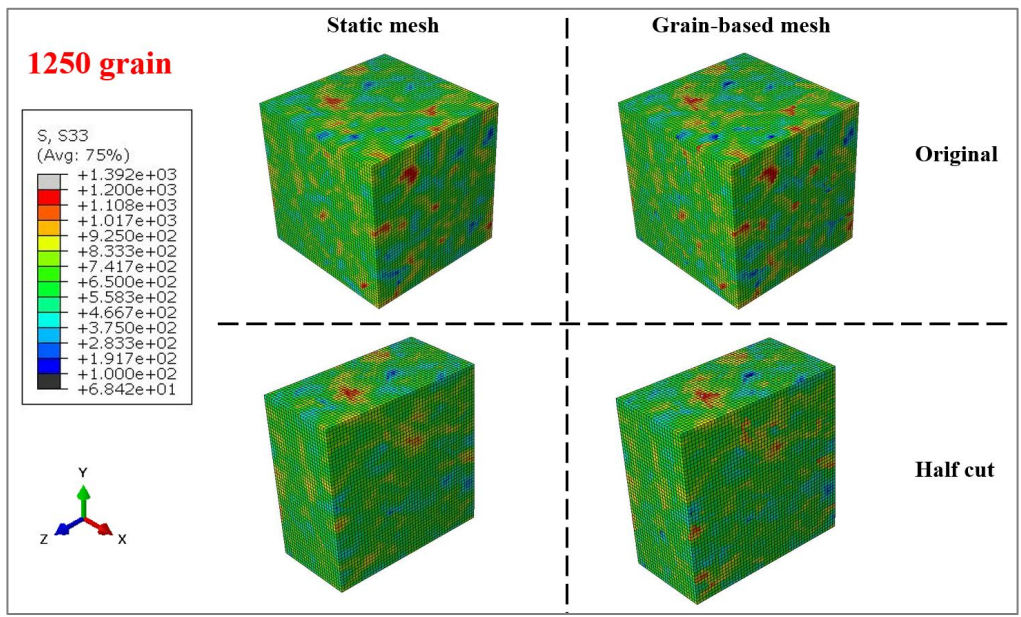

**Figure 14**. Original and half-cut contour maps for RVE with 1250 grains show the stress in the loading direction after tension to a total strain of 1.55%.

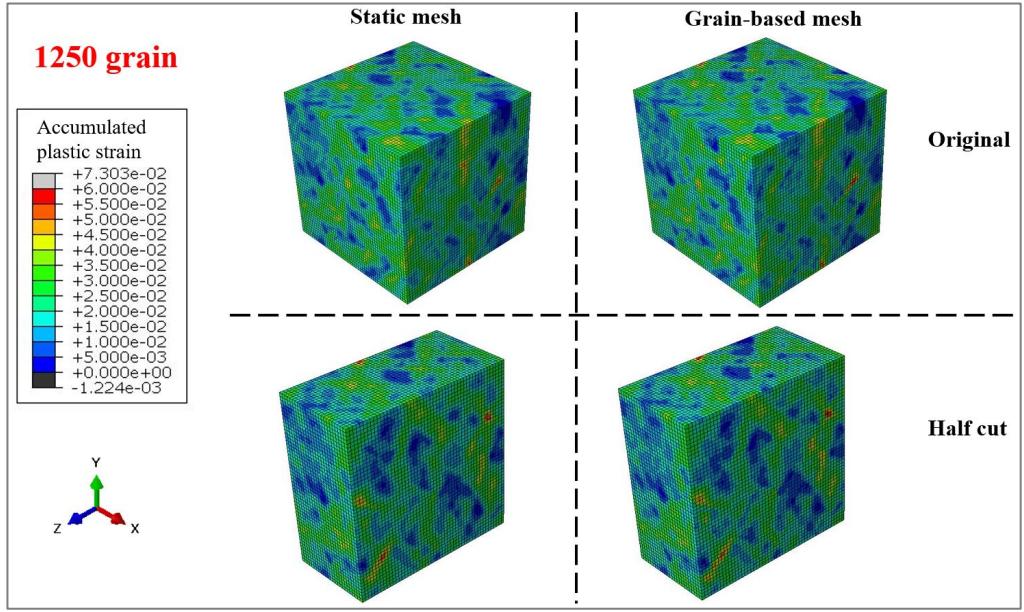

**Figure 15**. Original and half-cut contour maps for RVE with 1250 grains show accumulated plastic strain after tension to a total strain of 1.55%.

To evaluate the quantitative difference between the two discretization methods, the deviation measure is defined as

$$\text{Dev (\%)} = \frac{|\sigma_{static} - \sigma_{grain}|}{\sigma_{static}} \qquad (10)$$

Where $\sigma_{static}$ and $\sigma_{grain}$ are the stress obtained from realizations with static mesh and grain-based mesh, respectively. The deviation can be either macroscopic (global) when $\sigma_{static}$ and $\sigma_{grain}$ are the overall response of the RVE, or microscopic (local) when the two stresses are the local stress at the corresponding integration point. As illustrated in Figure 16, the macroscopic deviations at 1.55% total strain are relatively small, and they are all within 0.5%. The macroscopic deviations increase significantly when the number of grains within the RVE increases from 34 to 270, and the largest amount of deviation is developed for the RVE with the largest number of grains. A plateau is observed and there is no apparent increase in macroscopic deviations when the grain number further increases above 270.

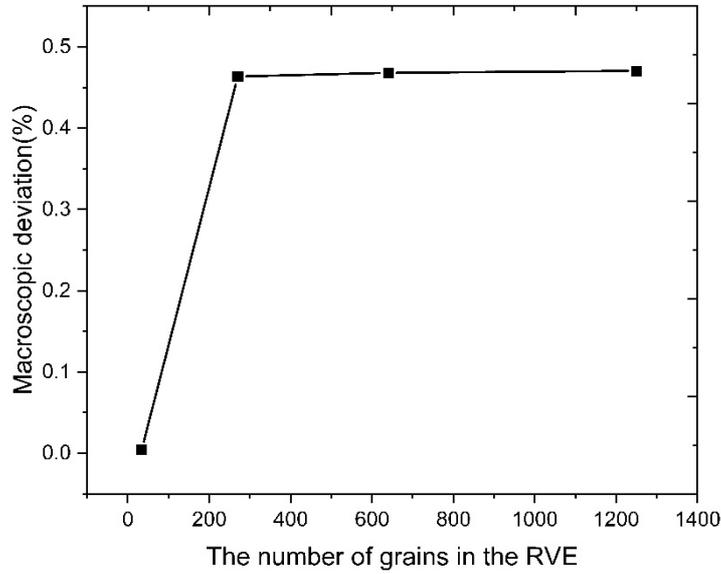

**Figure 16.** The macroscopic deviation calculated using Equation 10 for tensile loading to 1.55% strain.

At each integration point, the microscopic deviation can be evaluated by the local stresses obtained from RVE with static mesh and its corresponding RVE with grain-based mesh. Figure 17 compares the probability density of microscopic deviations calculated from RVEs with a different number of grains after 1.55% total strain. Interestingly, no apparent deviation was observed for the distribution of local deviations as the number of grains within an RVE increases. The majority of the microscopic deviations are less than 10%. Compared with macroscopic deviations, the microscopic deviations are more significant, and their value can range between 0-50%.

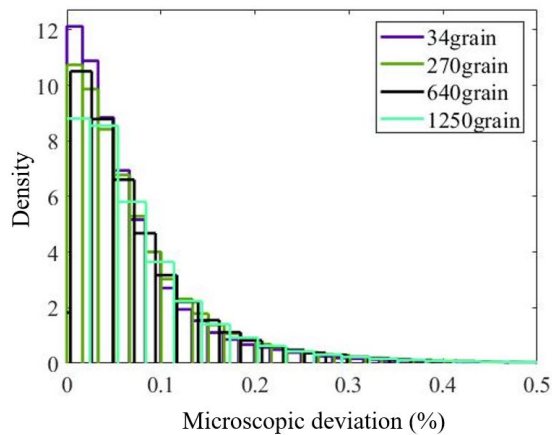

**Figure 17.** The microscopic deviations at all integration points for different numbers of grains within an RVE.

## 3.3 CPU time

To compare the computational efficiency of the CPFEM simulations with static and grain-based mesh, All the simulations were performed using a 48-core Intel Xeon Platinum 8268 computational node. A nonlinear implicit solver was used in Abaqus and the initial step, minimum and maximum step time for all simulations were 0.02s, 0.0001s and 1s, respectively. Table 3 lists the number of increments and total CPU time for all the simulations to achieve 1.55% strain. It is clear from Table 3 that static mesh reduces 5%-15% computational time compared to grain-based mesh, which suggests that static mesh shows a faster convergence rate than grain-based mesh. As reported in Section 3.1, more strain/stress discontinuities are observed in the RVE with grain-based mesh. More simulation time and increments are needed when there are stress discontinuities, while the static mesh reduces the simulation time by smoothing the discontinuities without introducing any apparent deviation in the macroscopic response.

Table 3. Total CPU time and the number of increments for all simulations.

| Grains per RVE | Total CPU time in hours | | The number of increments | |
|---|---|---|---|---|
| | Static mesh | Grain-based mesh | Static mesh | Grain-based mesh |
| 34 | 7.2 | 8.2 | 347 | 445 |
| 270 | 129.6 | 144.5 | 595 | 661 |
| 640 | 469.9 | 547.7 | 697 | 806 |
| 1250 | 1710.2 | 1781.8 | 938 | 975 |

## 3.4 Discussion

The present work aims to reveal the macroscopic and microscopic deviations in polycrystalline CPFEM with two discretization methods and different numbers of grains within an RVE. To the best of the authors' knowledge, this is the first research study that presents a detailed quantitative big data comparison of the simulation accuracy and computational time as a function of discretization methods, i.e. static mesh vs grain-based mesh. This research is essential when predicting the evolutions of local strain and stress fields during plastic deformation, especially while reproducing crack formation and growth. Such phenomena are challenging as the accuracy of these models strongly relies on the explicit meshing of grain structure. Both static mesh and grain-based mesh can well predict the location of strain/stress location and their results are consistent with each other. Although static mesh and grain-based mesh can reproduce almost identical stress-strain curves within a 0.5% difference, the local strain/stress response of the two discretization methods shows a large deviation in magnitude. The microscopic deviation can be as large as 50%, which is a significant and unreliable value for many service conditions.

As for the computational time, static mesh shows faster convergence and reduces 5%-15% computational time compared to grain-based mesh. Another advantage of static mesh is that it can be used to research multiple realizations without changing the RVE model once the model is validated. The only modification needed for static mesh is to provide a different set of seeds position and their corresponding crystal orientation. However, for grain-based mesh, each realization has its unique RVE model, and the element sets for each grain need to be assigned in the pre-processing stage. Because static mesh reduces both pre-processing and simulation time and provides comparable stress-strain curves to grain-based mesh, our results suggest that static mesh is recommended for CPFEM when calibrating the material parameters or when the macroscopic mechanical response is of interest. Nevertheless, the predicted deviations for the microscopic response are much more significant than the macroscopic response. When the local mechanical responses such as fatigue indicator parameter and accumulated plastic strain, are the main research interest, it is recommended to employ grain-based mesh to get improved solution accuracy as it can explicitly describe the material discontinuity at a proper level (element).

## 4. Conclusion

A large number of crystal plasticity finite element simulations were performed to perform a big data analysis of the effect of discretization methods and grain numbers on the solution accuracy of the predicted strain/stress filed. The grain structure in RVEs was discretized into the static mesh and grain-based mesh. Static mesh is integration point-based and the IPs in the same element can belong to multiple grains, while the minimum discretization unit for grain-based mesh is an element and each element can only belong to a single grain. Multiple sets of RVEs were subjected to displacement-controlled loading under periodic boundary conditions and all the simulations were analyzed by Abaqus using the UMAT subroutine. RVEs with static and grain-based mesh both can well predict the stress-strain curves and the stress/strain localization points in polycrystalline materials. A qualitative comparison of contour maps and stress profile reveals that more stress discontinuities exist in grain-based mesh compared to static mesh. When one element in static mesh belongs to two or more grains, it tends to smooth the stress profile at the grain boundary. Further investigation suggests that there is a large deviation in terms of the magnitude of local stress/strain predicted by the two discretization methods. The above findings remain the same when the number of grains within an RVE increases. A deviation parameter is defined to evaluate the macroscopic and microscopic differences caused by the static and grain-based mesh. The macroscopic deviations are relatively small and all within 0.5%, while the microscopic deviations are more significant and range between 0-50%. Static mesh shows faster convergence and reduces CPU time and pre-processing procedure compared to grain-based mesh. It is then suggested that static mesh is recommended when the macroscopic mechanical response is of interest and grain-based mesh is more desirable to choose when the microscopic behaviour of materials is the main research interest. This study provides useful guidance for discretization methods in CPFEM and should be applicable to other crystal structures.

## CRediT authorship contribution statement

**Jingwei Chen:** Conceptualization, Formal analysis, Validation, Visualization, Writing – original draft. **Zifan Wang**: Validation, Visualization. **Alexander M. Korsunsky:** Conceptualization, Methodology, Supervision, Formal analysis, Writing – review & editing.

## Data availability statement

The datasets used and/or analysed during the current study are available from the corresponding author on reasonable request.